\documentstyle[pre,aps]{revtex}
\input{psfig}
\draft
\begin{document}
\twocolumn[\hsize\textwidth\columnwidth\hsize\csname@twocolumnfalse\endcsname
\def\OSL{\mbox{{\bf OSL}}}
\def\DFL{\mbox{$\DF^L$}}
\def\DF{\mbox{{\bf DF}}}
\def\A{\mbox{{\bf A}}}
\def\x{\mbox{$\bf{x}$}}
\def\y{\mbox{$\bf{y}$}}
\def\max{\mbox{\mathrm{max}}}
\def\boldxi{\mbox{$\large \xi$}}
\def\boldeta{\mbox{$\large \eta$}}
\def\i{\mbox{$\mathrm{i}$}}
\def\Im{\mbox{$\mathrm{Im}$}}
\def\Re {\mbox{$\mathrm{Re}$}}
\include{epsf}
\title{Communication with Chaos over Band-Limited Channels}
\author{Nikolai F. Rulkov and Lev S. Tsimring}
\address{Institute for Nonlinear Science,
University of California, San Diego, La Jolla, CA 
92093-0402}
\date{\today}
\maketitle

\begin{abstract}
Methods of communications using chaotic signals use an 
ability of a chaos generator (encoder) and matched response 
system (decoder) to behave identically despite 
the instability of chaotic oscillations. 
Chaotic oscillations cover a wide spectral domain and can efficiently 
mask an information signal scrambled by  the chaotic encoder. 
At the same time the wide spectrum poses intrinsic difficulties 
in the chaotic decoding if the chaotic signal is transmitted 
over real communication channels with limited bandwidth. We 
address this problem both numerically and experimentally. Two 
alternative ways to improve communication with chaos over 
band-limited channels are investigated. The first method 
employs a matching filter in the decoder which compensates 
channel distortions of the transmitted signal.
This modification does not change the individual dynamics of 
chaotic systems in the synchronous state however the information
signal injected into the driving system, breaks the symmetry 
between encoder and decoder and therefore exact recovery is 
impossible. We show that this approach has limited ability 
for synchronization of chaotic encoder. The second approach
does not use adaptive compensation but relies on the design of chaotic
oscillators which produce narrow-band chaotic waveforms. 
\end{abstract}

\narrowtext
\vskip1pc]

\section{Introduction}

Nonlinear oscillators with chaotic dynamics are able to 
generate broadband non-periodic signals. Due to local 
instability of chaotic trajectories it is practically 
impossible to reproduce exact waveforms of the signals 
generated by a chaotic oscillator. However, it has been shown 
\cite{FY,pikovsky,AVR,volk89,PC90,winful,ogorzalek,roy} 
that two systems with perfectly matched parameters 
can exhibit stable regime of identical chaotic oscillations if 
adequately coupled. This property makes chaotic systems 
attractive for applications in private communications as an 
analog message can be hidden within a complicated 
structure of a chaotic waveform and still be reconstructed at 
the receiver using another nonlinear system whose parameters 
are closely matched to ones of the first system.\footnote{We do not 
address here an issue of digital data transmission based on chaotic
dynamics. Digital chaotic systems are essentially equivalent to 
stream cyphers known in cryptography\cite{crypt}} 

Numerous algorithms of chaotic encoders and decoders based on 
the stability of regimes of identical chaotic oscillations 
were suggested in the past several years. In most of these 
algorithms, the information signal $m(t)$ is added to 
the chaotic output $x_e(t)$ generated by a chaotic encoder
whose oscillations do not depend on $m(t)$. The mixture 
$x_e(t)+m(t)$ is transmitted to the decoder where it is used as a 
driving signal for the matched response system. Various implementations 
of the matched response systems were proposed, see for 
example~\cite{CP,cuomo,kocarev92,kennedy,murali,yu} 
and references therein. The common shortcoming 
of such methods of communication is that the driving signal which 
is ``distorted" by the message $m(t)$, does not perfectly fit 
the decoder. As a result, recovered message $m_r(t)$ will 
always contain some traces of chaotic waveforms no matter 
how perfectly the parameters of the decoder match those of the encoder. 
The amplitude of the chaotic contamination is small only if 
$m(t)$ is small as compared to the chaotic signal $x(t)$. Unfortunately, 
this makes the whole system susceptible to noise. 

A different approach to the problem of chaotic encoding and 
decoding was suggested in a number of 
papers~\cite{volk93,itoh,frey,chua,parlitz,feldmann,kocarev96}. The 
main idea of this approach is that information signal $m(t)$ 
is injected into the feedback loop of the chaotic encoder. As 
a result the distorted chaotic signal $x_e(t)+m(t)$ 
returns back to the generator and influences the dynamics of the 
encoder. Now when the signal $x_e(t)+m(t)$ is applied to the decoder, 
it generates oscillations of the response system which 
are identical to the oscillations in the encoding generator. 
The message $m(t)$ can be recovered using the open feedback 
loop of the response system. In this case (in the absence of 
noise), after initial transients, the information can be restored 
exactly. To be more specific consider the chaotic encoder of the form 
\begin{eqnarray}
\frac{dx_e}{dt}&=&f(x_e,{\bf y}_e,u_e),\nonumber\\
\frac{d{\bf y}_e}{dt}&=&{\bf F}(x_e,{\bf y}_e,u_e),\nonumber\\
u_e&=&x_e+m(t),
\label{encoder}
\end{eqnarray}
where the scalar $x_e$ and the vector ${\bf y}_e$ are the 
dynamical variables of the chaotic generator, $m(t)$ is the 
information signal and $u_e$ is the output of the 
encoder. It is assumed that the choice of the vector function 
$(f,{\bf F}^T)^T$ provides the chaotic behavior of the system 
(\ref{encoder}). The dynamics of the decoder is described by 
the similar equations
\begin{eqnarray}
\frac{dx_d}{dt}&=&f(x_d,{\bf y}_d,u_d),\nonumber\\
\frac{d{\bf y}_d}{dt}&=&{\bf F}(x_d,{\bf y}_d,u_d),\nonumber\\
m_r(t)&=&u_d(t)-x_d,
\label{decoder}
\end{eqnarray}
where $u_d(t)$ is the external signal driving the decoder. Evidently, 
if $u_d(t)\equiv u_e(t)$, solution $x_d = x_e,\ {\bf y}_d=
{\bf y}_e,\ m_r= m$ satisfies (\ref{decoder}).
Stability of this regime of identical chaotic oscillations 
in systems (\ref{encoder}) and (\ref{decoder}) can be 
assured by the proper choice of the vector function $(f,{\bf F}^T)^T$. 
Then starting from arbitrary initial conditions,
$x_e(t)-x_d(t) \rightarrow 0$ as $t \rightarrow \infty$, and 
therefore, $m_r(t) \rightarrow m(t)$. 
The mathematical background for stability of regimes of identical 
chaotic oscillations can be found elsewhere~\cite{ashwin}.
We would like to note that the described method of encoding 
and decoding is conceptually equivalent to a well known method 
of data encryption by means of digital data scramblers, 
see~\cite{shift} and references therein. 

The fact that the chaotic signal is broadband, poses serious problems 
if it is to be transmitted through a band-limited channel. Due to
the nonlinear nature of the matched response system, if some part of the 
spectral components of the chaotic driving signal is not transmitted, 
or distorted, it affects the whole spectral domain of the 
signals produced in the response system. Even simple variations
of channel gain or pure phase distortions in the channel (so called 
all-pass filters) may spoil drastically the response behavior\cite{ocbf96}.

Several recent papers \cite{carroll94,carroll95,cp96,cyzw96,ocbf96} 
addressed this problem by attempting to eliminate the distortion 
imposed by the channel 
either using an inverse filter\cite{carroll94,ocbf96} or augmenting the 
distorted signal by the signal of the second system passed through 
a complimentary band-stop filter\cite{carroll95}. For a mild 
first-order filter simulating the channel distortion, synchronized chaotic 
oscillations were achieved, however with less degree of stability 
as compared with the no-filter case.  
Both of these methods assume that the channel impulse response function 
is known at the receiver.
In practice, however, channel characteristics are often unknown
{\it a priori} and may change in the course of the transmission.
\footnote{If the channel only changes the amplitude of the signal,
adaptive schemes of channel equalization can improve the quality of
chaotic synchronization without {\em a priori} knowledge of
the channel gain (see \cite{cyzw96,ocbf96}).}

When one wishes to apply synchronized chaotic signals for private
communication, the ultimate task is of course transmitting useful 
information without errors. Limited bandwidth of the channel 
complicates this task significantly and in fact makes channel 
equalization for such transmission even 
more difficult than just providing good chaotic synchronization
without information transmission. In particular, the method of
\cite{carroll95} will still produce significant errors in 
signal reconstruction unless the channel characteristic is known
{\em at the transmitter} and the information signal is pre-processed 
accordingly (see Section \ref{sec2} for details).

Our approach to using synchronized chaos for signal transmission 
is based on a different principle. We propose to communicate by 
means of a chaotic carrier which is wide enough (spectrally) to 
mask the information signal, but is still narrow enough to fit into 
the channel bandwidth without much distortion. In 
this paper we examine two such methods in numerical 
simulations and illustrate the second method by 
experiments with electronic circuits. 
The paper is organized as follows. In Section \ref{sec2} we review the 
methods for channel equalization (channel inversion and channel 
compensation) as applied to chaos synchronization.
In Section \ref{sec3} we discuss information transmission using 
channel compensation.
In Section \ref{sec4} we describe two alternative ways  
to cope with channel distortions which both rely on the design of 
chaotic systems such that only a narrow-band signal is being 
sent through a channel. In Section \ref{sec5} some results 
of experiments with electronic circuits are presented. 

\begin{figure}
\centerline{\psfig{figure=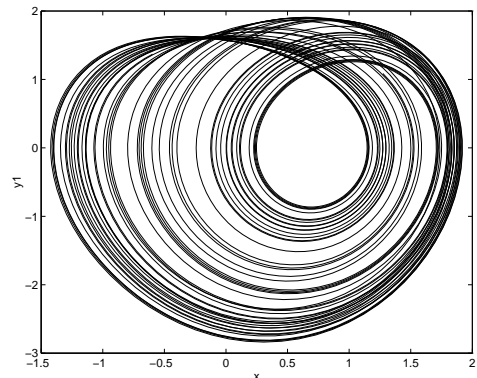,height=2.0in}}
\caption{ Attractor of the driving system (\protect\ref{encoder}) with
rhs of (\protect\ref{system}),(\protect\ref{feigen}) in the
$(x,y_1)$-plane
\label{fig1}
}
\end{figure}

\begin{figure}
\centerline{\psfig{figure=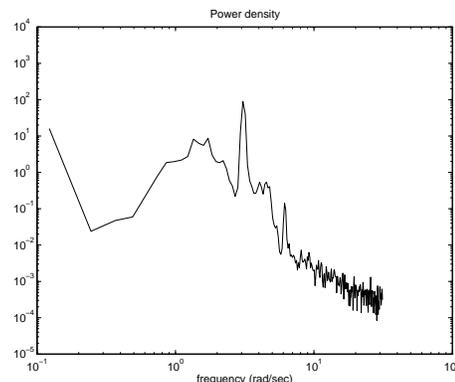,height=2.0in}}
\caption{ Power spectrum of the $x$ component of system
(\protect\ref{encoder}),(\protect\ref{system}),(\protect\ref{feigen})
\label{fig2}
}
\end{figure}

\section{Channel equalization for synchronization of chaos}
\label{sec2}

\begin{figure}
\centerline{\psfig{figure=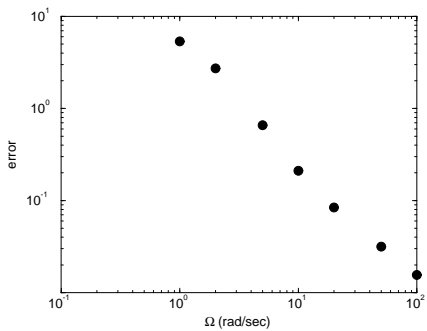,height=1.9in}}
\vspace{-0.0in}
\caption{ Rms synchronization error $x_d-x_e$ for
system (\protect\ref{encoder})-(\protect\ref{feigen})
as a function of the cutoff frequency of the channel.
\label{fig3}
}
\end{figure}

Standard design of  a system of two synchronized
chaotic oscillators usually assumes a perfect
channel with infinite bandwidth and no phase distortions (see, e.g.
\cite{cuomo,kocarev96}). However even a mild filter placed between 
the oscillators drastically impairs synchronization\cite{cp96}.   
We illustrate this problem by simulating a transmitter and 
receiver in the form of (\ref{encoder}),(\ref{decoder})  
with ${\bf y}=[y_1,y_2],\ {\bf F}=[F_1,F_2]$, where
\begin{eqnarray}
f(x,{\bf y},u)&=&y_1,\nonumber\\
F_1(x,{\bf y},u)&=&y_2,\nonumber\\
F_2(x,{\bf y},u)&=&-ay_2 -by_1-cx+dG(u).
\label{system}
\end{eqnarray}
Nonlinearity is taken in the form
\begin{eqnarray}
G(u)&=&\left\{\begin{array}{l} 1-u^2,\ \mbox{at}\ -0.95<u<2,\\
0.0025,\ \mbox{at}\ u<-0.95,\\
-3,\ \mbox{at}\ u>2,
\end{array}\right.
\label{feigen}
\end{eqnarray} 
and parameters $a=0.73,b=2.63,c=0.2,d=2.0$.
For the driving system $u_e=x$ (no information is being transmitted), and
for the response system 
\begin{equation}
u_d(t)=c(t)\otimes u_e(t),
\label{filter}
\end{equation}
where $c(t)=\Omega_c^{-1}\exp(-\Omega_c t)$ is an impulse response function 
of the channel, and signal $\otimes$ denotes linear convolution.
The chaotic attractor for the system (\ref{encoder}),(\ref{system}),(\ref{feigen}) on $(x,y_1)$-plane 
is shown in Figure \ref{fig1}. The power spectrum of the $x$-component 
of the driving system is shown in Figure \ref{fig2}. In Fig.\ref{fig3} 
we show the rms synchronization error $||x_e-x_d||d$  as a function of 
the cutoff frequency $\Omega_c$.
Comparing Figs.2 and 3 one can see that the synchronization 
errors accumulate quickly and become significant when most of the 
power spectrum of the chaotic signal is still well below $\Omega_c$.

\begin{figure}
\centerline{\psfig{figure=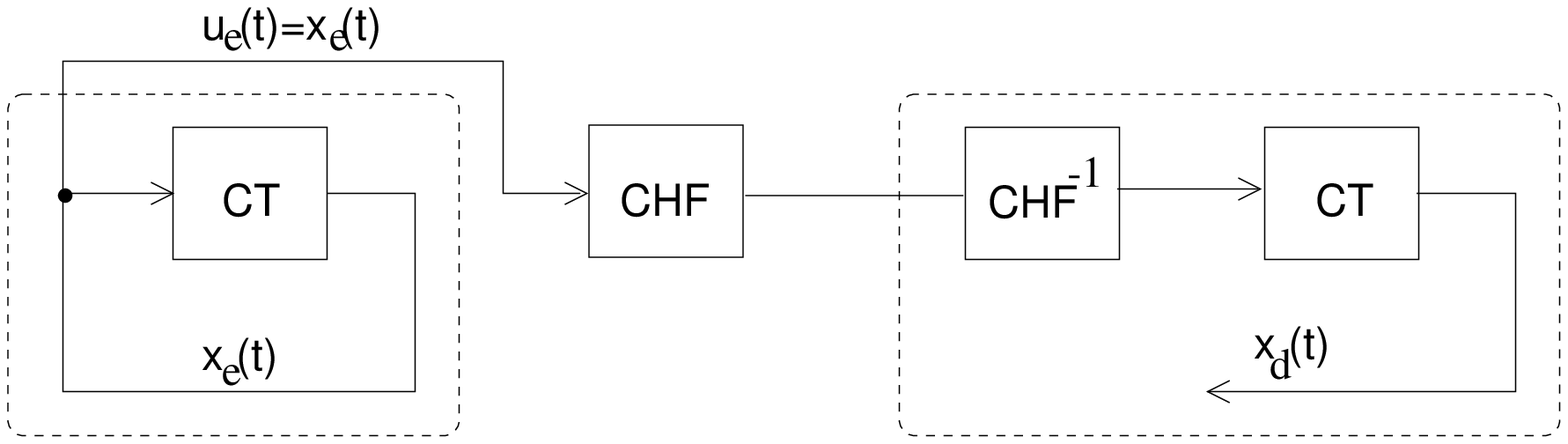,height=0.95in}}
\vspace{0.2in}
\centerline{\psfig{figure=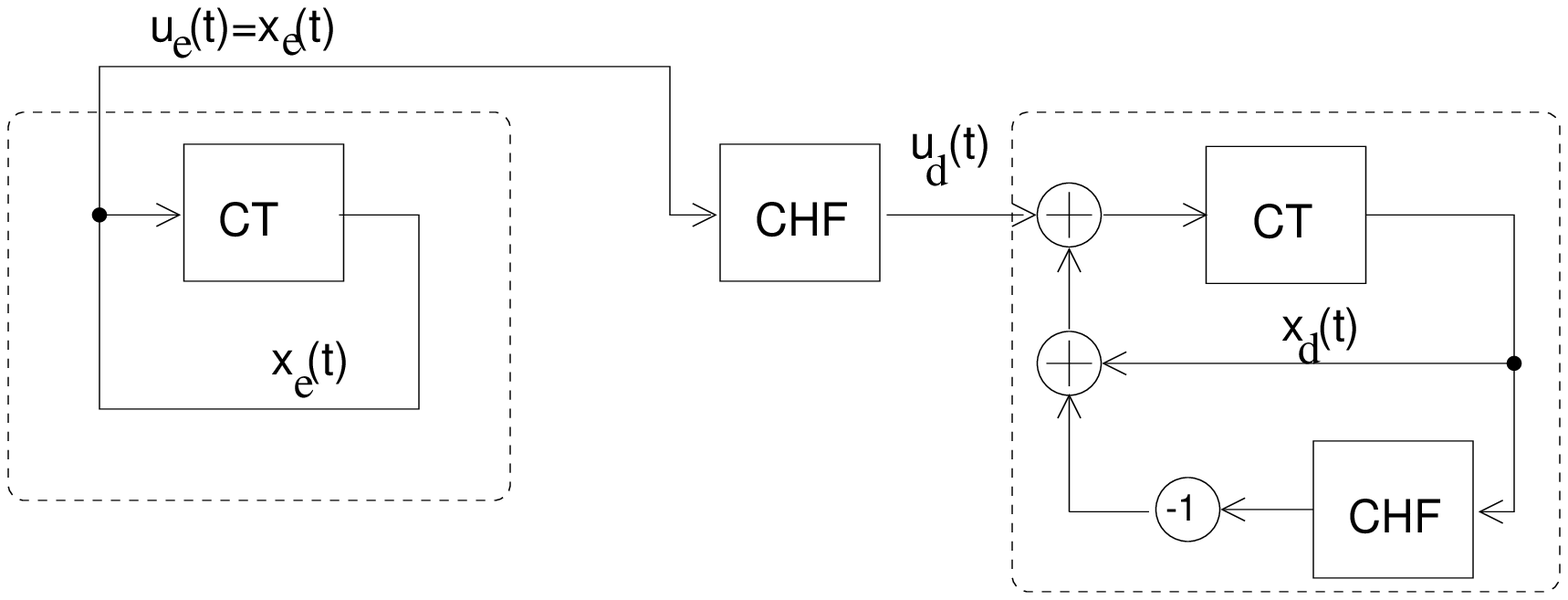,height=1.3in}}
\caption{{\em (a)} Block-diagram of channel equalization method
based on filter inversion\protect\cite{carroll94};
{\em (b)} Block-diagram of channel equalization method
based on complimentary filtering in the feedback loop of the
response system\protect\cite{carroll95};
\label{fig4}
}
\end{figure}

Two different approaches have been proposed to cope with 
channel distortions in order to preserve synchronization 
of chaotic oscillators. Both these methods have an advantage 
that ideally they allow for exact synchronization without 
affecting the structure of chaotic attractors. The first scheme 
\cite{carroll94,cyzw96,ocbf96} involves
learning the channel response function $c(t)$ and building an inverse filter,
$c^{-1}(t)$ (see Figure \ref{fig4}a).   Apparently, this 
method requires that channel distortions are invertible, and the 
inverse filter is stable, which
in not necessarily true in general.  It also is very sensitive to noise 
in the channel. Indeed in the presence of noise the ``restored'' signal 
will have a form $u'=c^{-1}(t)\otimes (c(t)\otimes u(t) +\eta(t))$ 
($\eta(t)$ denotes random noise), and so the noise outside the passband 
of the filter $c(t)$ will be amplified by the inverse filter $c^{-1}(t)$.

The second method due to Carroll\cite{carroll95} employs augmenting
the chaotic signal filtered be the channel, by a signal of the
decoder passed through a complimentary band-stop filter 
\begin{equation}
u_d(t)=c(t)\otimes u_e(t) + (1-c(t))\otimes x_d(t) 
\label{augment}
\end{equation}
(see Figure \ref{fig4}b). In this case the filter response function also must 
be learned at the receiving end. As compared to the previous method, 
there is no filter inversion and associated with it problems of filter
instability and noise amplification. As Carroll\cite{carroll95} has shown
for a different system, for mild channel distortions (first-order filter) 
exact chaos synchronization can be achieved with this
type of driving, however the conditional Lyapunov exponent characterizing
stability of the synchronized state is less (by magnitude) than in the 
no-filter case.
Clearly, it is the last term in (\ref{augment}) which provides a
positive feedback in the response system and causes the reduction
of stability. Now, the question remains of how one can use
this method for transmitting information masked by the chaotic 
signal. Carroll\cite{carroll95} proposed simply to 
add an information component to the chaotic signal as
in \cite{cuomo}, and recover the information by 
observing the synchronization error at the response system. However this 
method does not yield good signal recovery even for a prefect channel, 
and cannot be used for high-quality transmission. 
In the next section we discuss the possibility of combining Carroll's 
method \cite{carroll95} with a more sophisticated 
scheme~\cite{kocarev96,feldmann} of
signal transmission. 

\section{Communication with chaotic signals - channel compensation}
\label{sec3}

\begin{figure}
\centerline{\psfig{figure=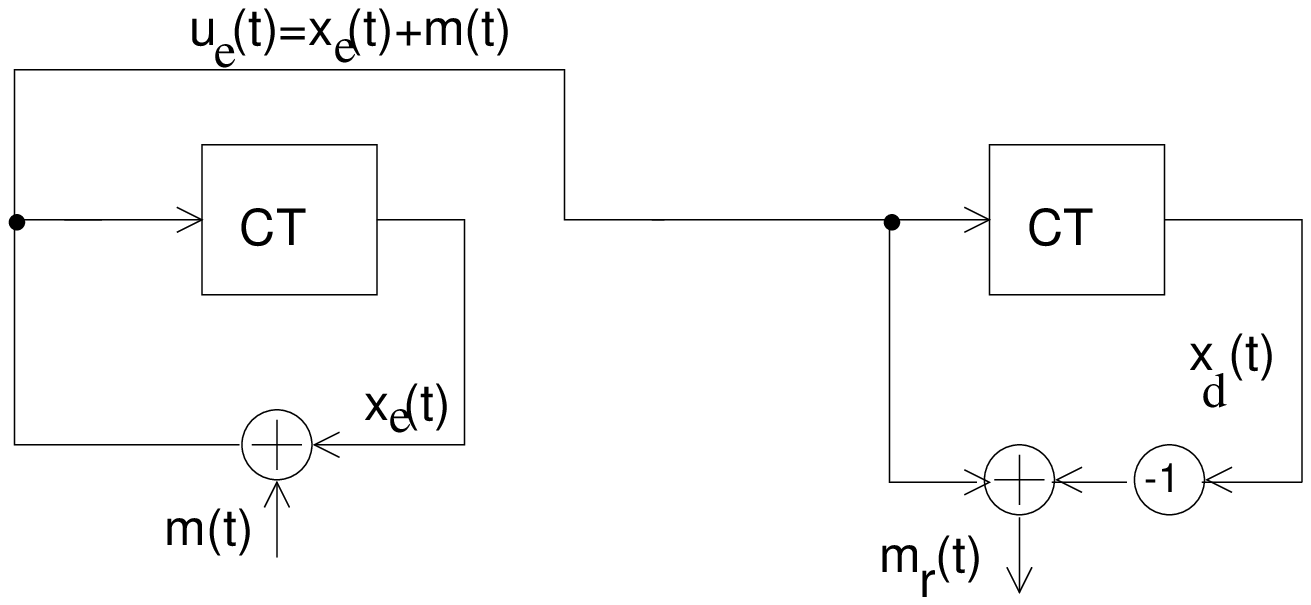,height=1.4in}}
\vspace{0.2in}
\centerline{\psfig{figure=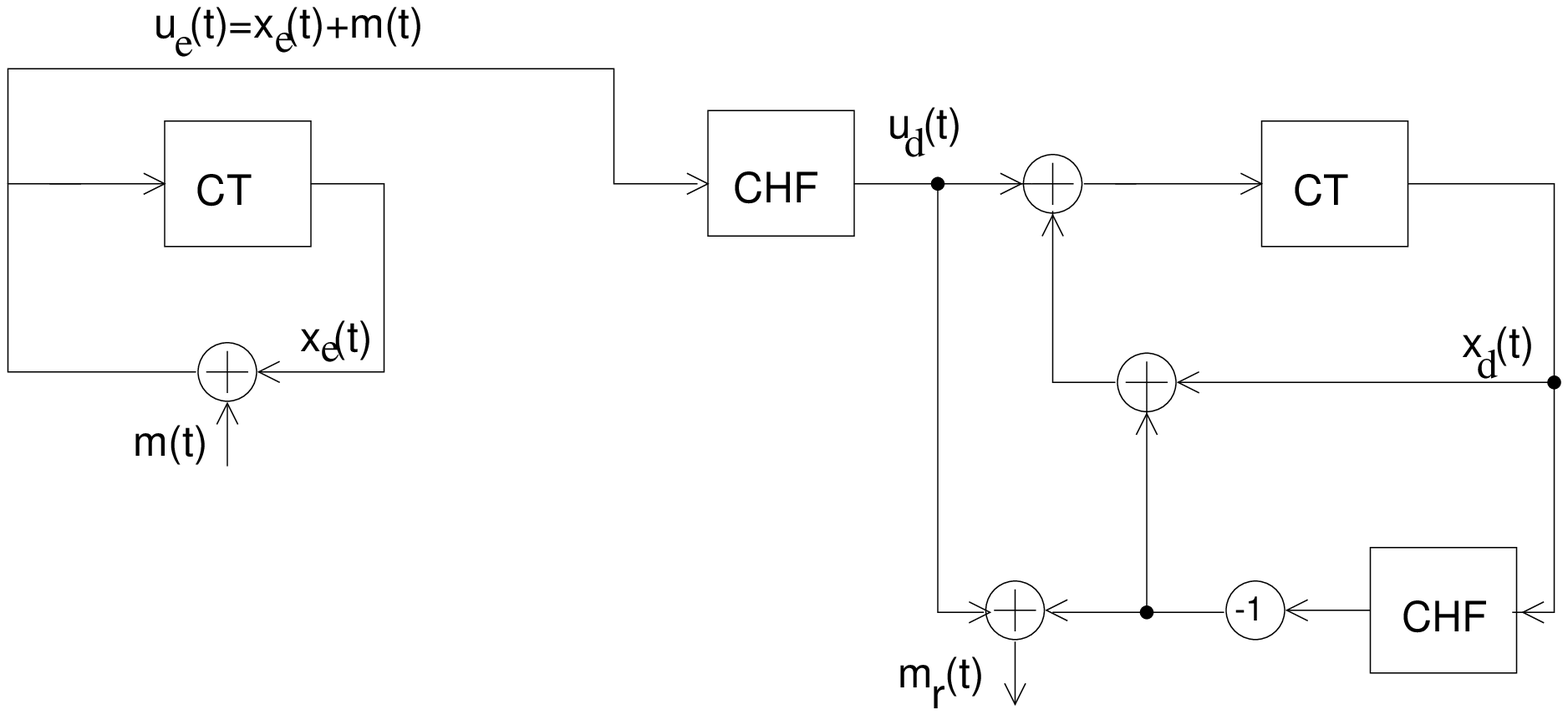,height=1.4in}}
\vspace{0.2in}
\centerline{\psfig{figure=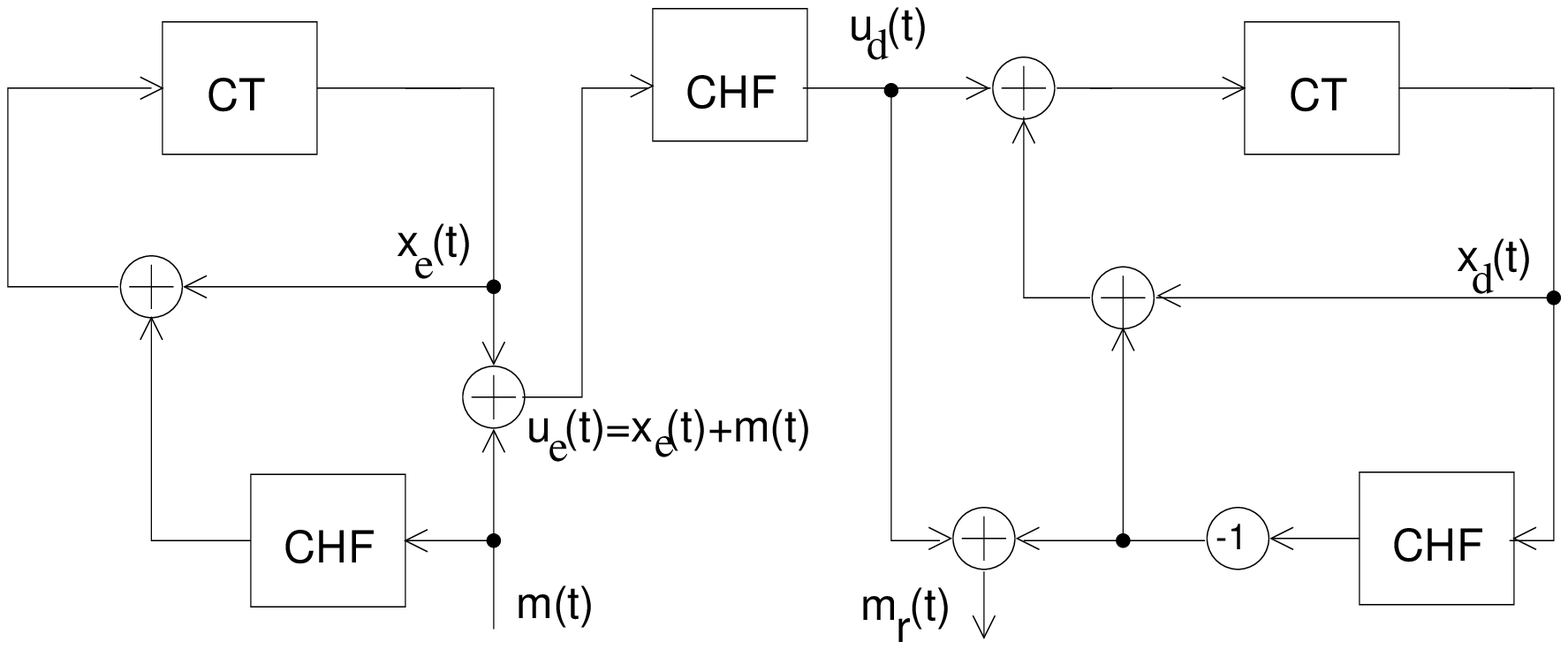,height=1.3in}}
\vspace{0.2in}
\caption{
{\em (a)}Block-diagram of the standard signal transmission
with chaos\protect\cite{volk93};
{\em (b)} Standard scheme combined with the channel equalization method
based on complimentary filtering in the feedback loop of the
response system;
{\em (c)} Injecting the information signal filtered with a filter
matching the channel filter, restores the symmetry between encoder and
decoder.
\label{fig5}
}
\end{figure}

\begin{figure}
\centerline{\psfig{figure=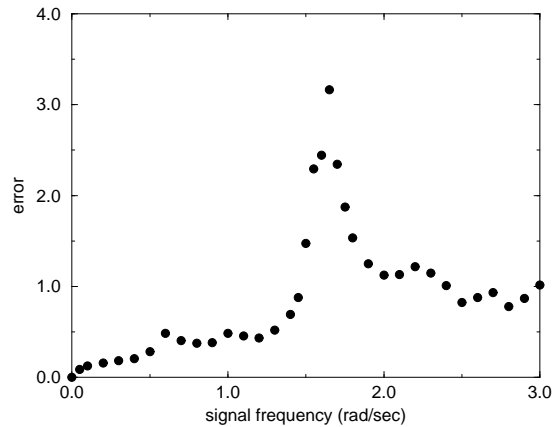,height=2.5in}}
\vspace{0.0in}
\caption{
Relative rms signal recovery error $||m_r-m||$ for sine wave
$m(t)=0.1\sin(\omega t)$ as a function of its frequency $\omega$
for information transmission using
system (\protect\ref{encoder})-
(\protect\ref{feigen}) and the method of Fig.\protect\ref{fig5}a
\label{fig6}
}
\end{figure}

In this section we consider information transmission using chaotic 
masking in the presense of channel distortions. 
The original method \cite{volk93} of signal transmission with chaos is
illustrated by Figure 5a. The message $m(t)$ is injected into the
feedback
loop of the driving system and is recovered at the decoder by
subtracting the
output of the chaotic transformer (CT) $x_d(t)$ from the transmitted signal
$u_d(t)$.
Without channel distortions this method yields exact recovery of the
message after an initial transient.
Let us first try to apply this method to the scheme of Fig.\ref{fig4}b
(see
Figure~\ref{fig5}b).
At the response system, the message is recovered by subtracting
the filtered feedback signal $x_d$ from the received signal
$u_d\equiv c(t)\otimes(x_e(t)+m(t))$ .  Unfortunately, when channel
distortions are present, recovered signal remains corrupted because the
symmetry between the encoder and decoder is broken. The decoder
receives a sum of the filtered chaotic signal with the {\em filtered}
message, whereas the encoder 
is fed by the {\em unfiltered} message. We illustrate this
asymmetry
by attempting to transmit a sine wave $m(t)=0.1\sin(\omega t)$
from the driving to the response system (\ref{system}),(\ref{feigen}) 
with the first-order filter (\ref{filter}) with a 
cutoff frequency $\Omega_c=5$ rad/sec simulating the communication channel. 
In Figure \ref{fig6} the rms recovery error relative to the standard 
deviation of the information signal is shown as 
a function of the signal frequency $\omega$. For small $\omega$, the filter 
does not affect the information signal and so the quality of recovery is 
high. However, for higher frequencies (which are still within the passband of
the channel) the filter starts to affect the 
amplitude as well as the phase of the transmitted information signal  and
recovery error grows rapidly. Moreover, the distorted information signal
can even break synchronization between the transmitter and receiver. The
large peak near 1.6 rad/sec is due to a resonance between the sinusoidal 
signal and the chaotic oscillator.

The only possible way to provide an exact synchronization when the 
information component is present is to learn the channel transfer function 
{\em at the transmitter} and, using a matched filter, pre-process 
the information signal accordingly before injecting it into
a feedback loop of the encoder. The block-diagram of this modification
is shown in Figure \ref{fig5}c. The decoding procedure which is the same as in 
Figure \ref{fig5}b, yields a {\em filtered} message $c(t)\otimes m(t)$ 
or exactly the same signal as one could possibly receive without mixing 
it with a chaotic component. Numerical simulations show that indeed the
recovery error remains within the accuracy of the integrator (10$^{-10}$
for signal frequencies throughout the passband of the channel). 
Thus, pre-processing of the information
signal cures the problem of asymmetry in the encoder-decoder and yields
exact synchronization and signal recovery. However, as was noted before,
this method does not prevent the instability which may occur due to
the compensating signal in the feedback loop of the response system.

\section{Communication with chaotic signals -- narrow-band chaotic 
system design} \label{sec4}

\begin{figure}
\centerline{\psfig{figure=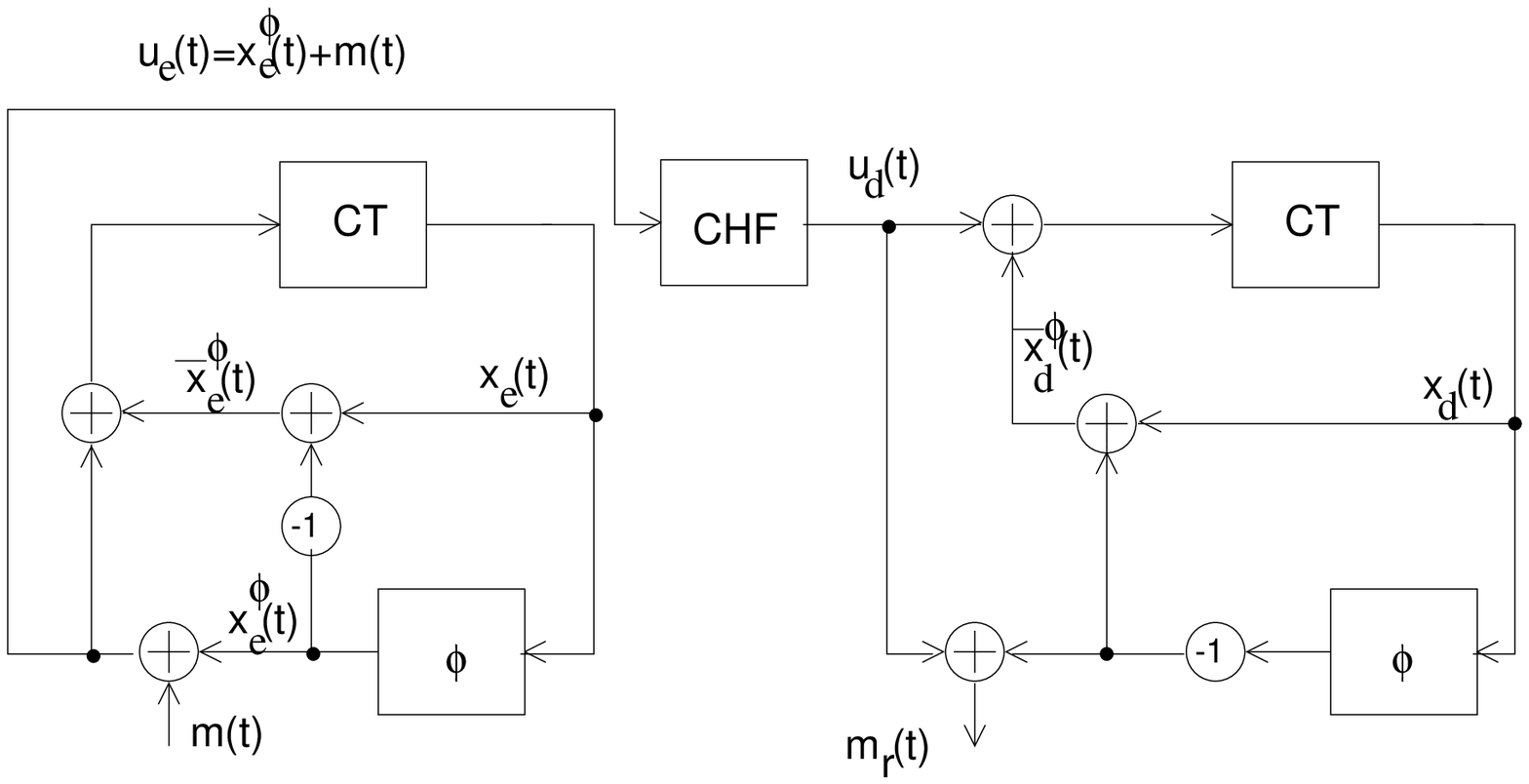,height=1.5in}}
\vspace{0.2in}
\centerline{\psfig{figure=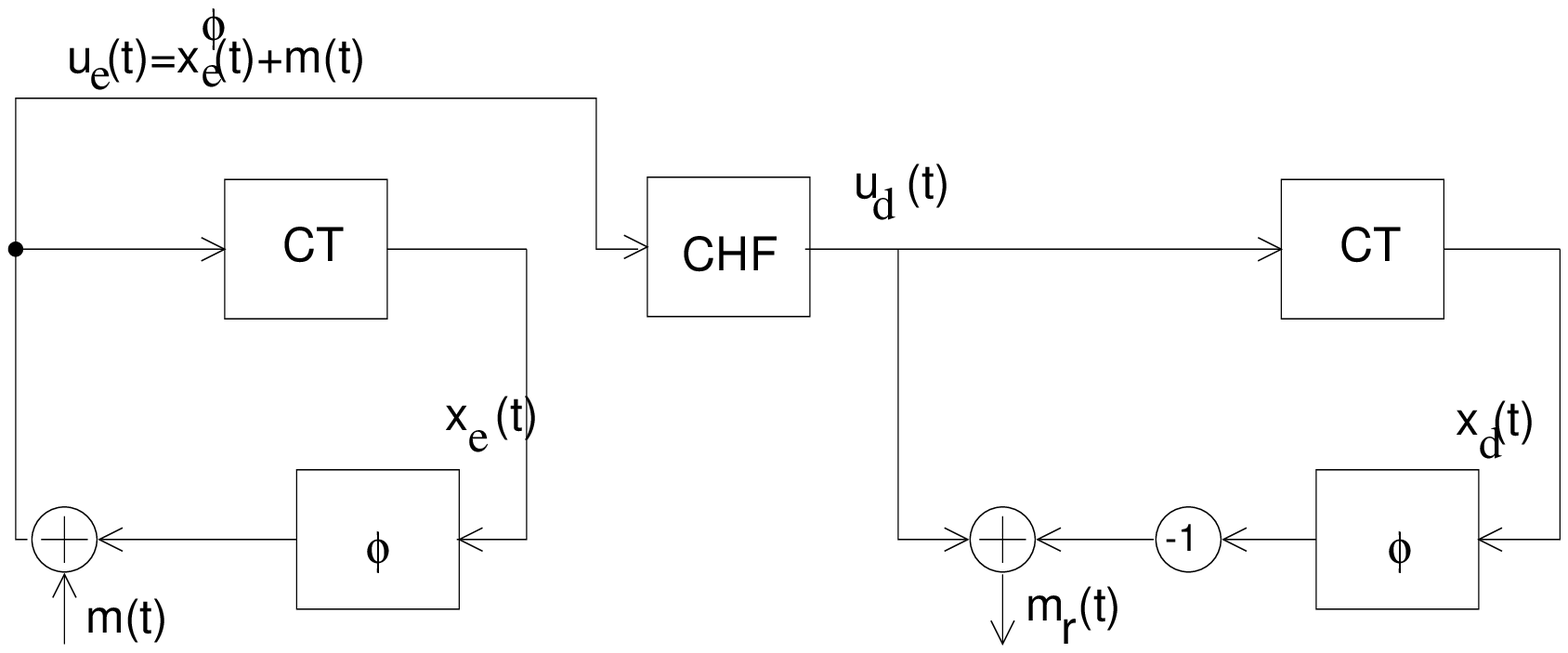,height=1.3in}}
\vspace{0.2in}
\caption{
{\em (a)} Block-diagram of information transmission with
the narrow-band chaotic masking signal
based on complimentary filtering in the closed feedback loops of
both driving and response systems (NBCL);
{\em (b)} Same for the open-loop response system  (NBOL).
\label{fig7}
}
\end{figure}

In the previous section we described a method of communication with
chatic signals over a band-limited channel based on matching the system 
properties with the properties of the channel.  The advantage of this 
method is that it preserves the structure of the original chaotic attractor 
for arbitrary channel response functions.  Unfortunately, this method does
not guarantee the stability of chaotic synchronization due to the
feedback in the second system. Moreover, this kind of channel compensation 
is difficult to achieve in practice, when channel parameters vary 
over time  or the same encoded message is to be transmitted via different
channels simultaneously. 

A different approach to achieve the same goal is to prepare a rather 
narrow-band chaotic signal which would not be significantly 
affected by the channel distortions. If, in addition, the 
information message is also narrow-band, it can be recovered 
at the decoder without distortion. In this section we describe
two methods of creating narrow-band chaotic signals. 

\begin{figure}
\centerline{\psfig{figure=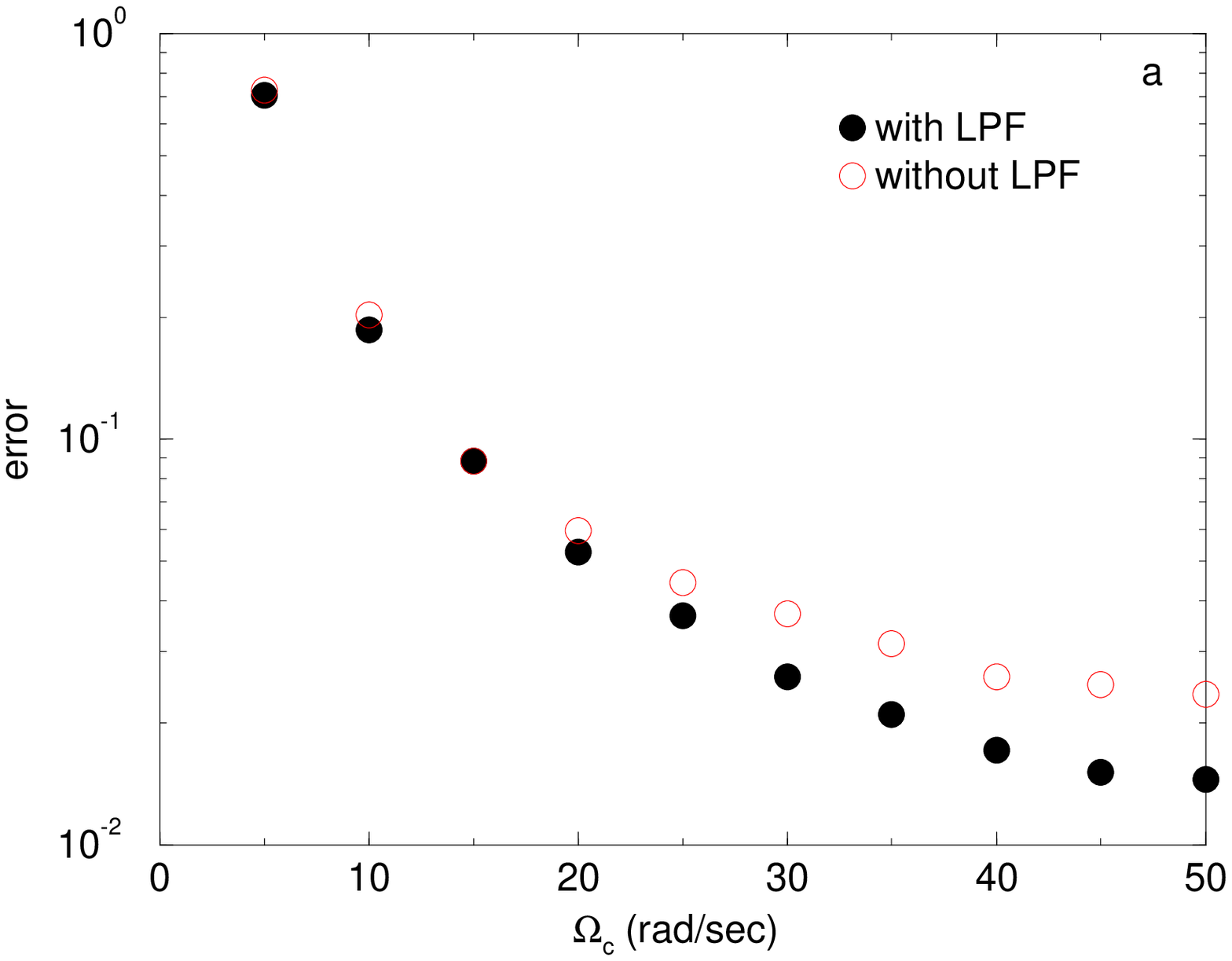,height=1.5in}}
\vspace{0.0in}
\centerline{\psfig{figure=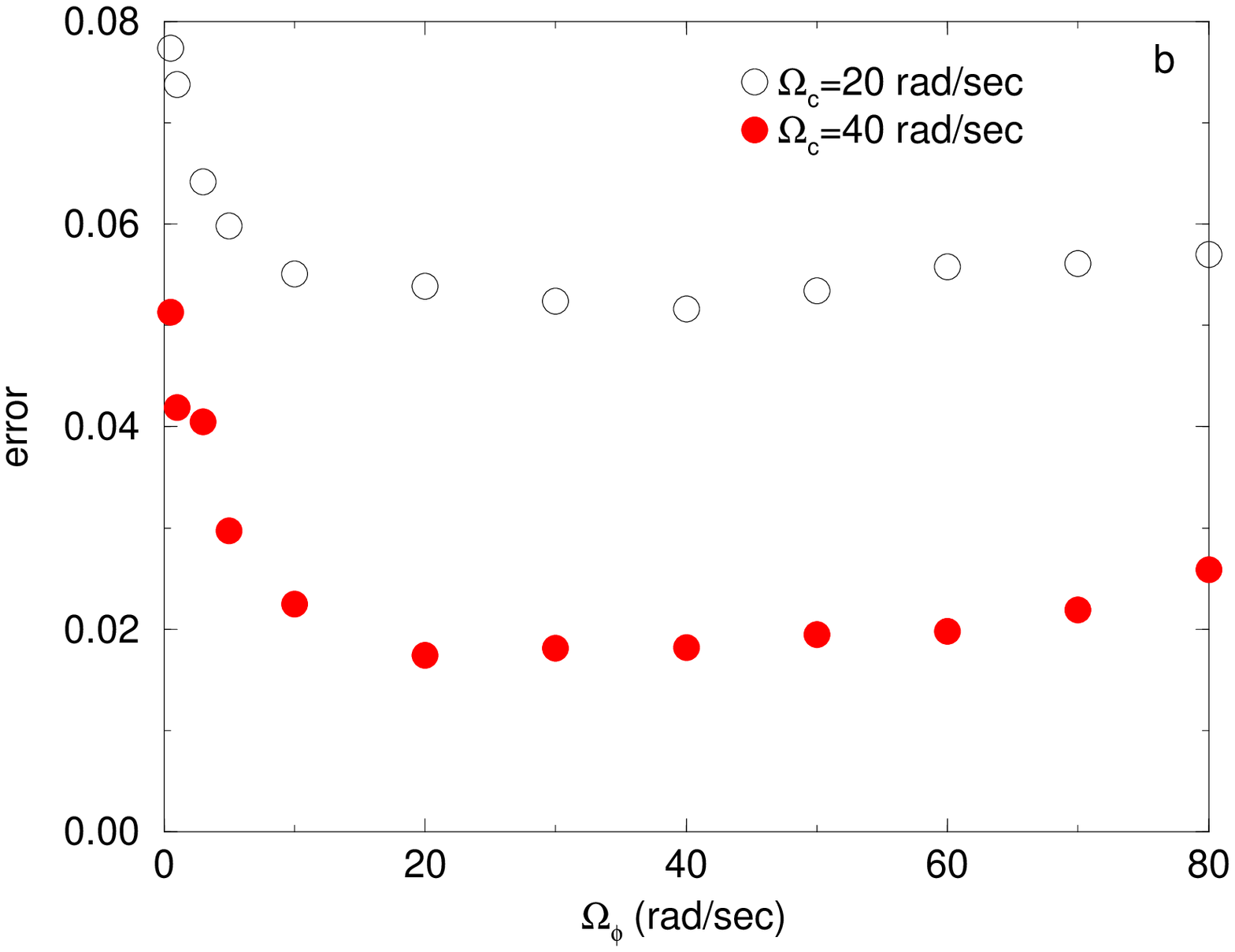,height=1.5in}}
\vspace{0.0in}
\caption{{\em (a)}Relative rms signal recovery error for sine wave
$m(t)=0.1\sin(\omega t)$ as a function of the cutoff frequency of the
channel for the NBCL scheme Fig.\protect\ref{fig7}a with and without
LP filter in the feedback loop;
{\em (b)} same as  function of the cutoff frequency of the LP filters
for
two cutoff frequencies of the channel.
\label{fig8}
}
\end{figure}

\subsection{Method 1}
The first method is shown schematically in Figure \ref{fig7}a. It is somewhat 
similar to 
the channel compensation method of Section \ref{sec2}, however 
the filters here are independent of the channel characteristics.
The $x$-component of the encoder is divided into a filtered part 
$x_e^\phi=\phi\otimes x_e$ and a complimentary part 
${\bar x}_e^\phi=x_e-x_e^\phi$. The information signal $m(t)$ is added to 
$x_e^\phi$ and then the sum $u_e=x_e^\phi+m(t)$ is sent to the decoder.
This sum is narrow-band since by assumption both the message 
and the filter $\phi$ are narrow-band. Further, the complimentary 
signal ${\bar x}_e^\phi$ is added to $u_d$. Therefore, the
signal which is injected into the nonlinear element of the system is
simply $x_e+m$, as if there were no filter in the feedback loop. 
At the decoder, the $x_d$-component is 
similarly split into the filtered part $x_d^\phi=\phi\otimes x_d$ 
and its complimentary part ${\bar x}_d^\phi=x_d-x_d^\phi$. Component
 ${\bar x}_d^\phi$ is added to the received signal $u_d$ and the sum is 
injected into the nonlinear element. The information signal is recovered
as a difference  $m_r(t)=u_d-x_d^\phi$. It is easy to see that
if the transmitted signal is passed through a channel without distortion
($u_d=u_e$), the message is recovered at the decoder exactly.
We call this scheme narrow-band, closed-loop design (NBCL).

One advantage of this scheme is that no matter how narrow the filter
$\phi$ is, the chaotic dynamics of the systems remains unchanged. Yet
the transmitted chaotic signal can be made arbitrarily narrow.
The disadvantage however is that this method again does not guarantee 
stable synchronization, since the component  ${\bar x}_d^\phi$ provides 
a closed feedback loop in the response system.   Clearly, the narrower 
filter  $\phi$ is, the more power is passed to ${\bar x}_d^\phi$ and
the synchronization becomes less stable.   Hence, the interplay
of these two counteracting factors  determines the quality of
transmission.

We applied this method to a chaotic encoder and decoder described by 
Eqs.(\ref{encoder}),(\ref{decoder}),(\ref{filter}). In the feedback 
loops of the oscillators we used a 4-th order low-pass elliptic filter 
with a stopband attenuation of 10 dB, passband gain of 0.6  and the cutoff 
frequency $\Omega_c$ which was adjusted in order to optimize the
performance of the transmission for a given channel. 
Figure 8a compares the rms error for transmission of the sine
wave $0.1\sin(\omega t)$ with $\omega=$0.5 rad/sec with the case when 
no filtration is applied to the transmitted chaotic signal. For relatively
high cutoff frequencies of the channel $\Omega_{c}$ pre-filtration yields 
certain improvements which however vanishe for lower $\Omega_{c}$. 
The reason 
for this is that it is impossible to reduce the cutoff frequency of
the LP filters below some value as the synchronization 
becomes unstable for low $\Omega_{\phi}$, and therefore the pre-filtration 
gives no improvement.  In Figure 8b this effect is illustrated by
the dependence of the rms error on the cutoff frequency of the LP
filters for two values of the cutoff frequency of the channel filter 
$\Omega_c$. The error initially drops with $\Omega_{\phi}$ due to reduction of
channel distortions, but then rises due to the instability 
of synchronization with respect to high-frequency components.

\subsection{Method 2}
\begin{figure}
\centerline{\psfig{figure=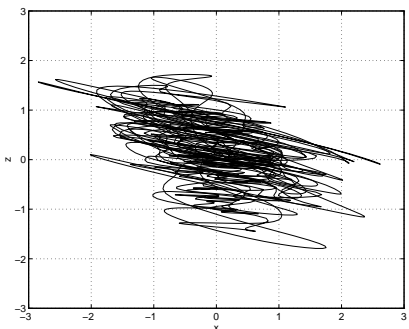,height=1.8in}}
\vspace{0.0in}
\caption{Chaotic attractor of the system (\protect\ref{encoder}),
(\protect\ref{system}),(\protect\ref{filter}) with nonlinearity
$G(u)=\sin(15u)$ and fourth-order Butterworth bandpass filter
(center frequency 1.5 rad/sec, banwidth 0.5 rad/sec) in the
feedback loop.
\label{fig9}
}
\end{figure}

As we mentioned above, method 1 (NBCL) guarantees chaotic behavior of
both driving and response systems for any filters used in their feedback
loops, but it does not guarantee the synchronization between encoder and
decoder.
It is desirable to couple two chaotic system by a narrow-band 
signal and still have a robust chaotic synchronization. This goal can be
achieved by changing the structure of chaotic oscillators in such a way 
that one of its internal variables is itself narrow-band, and therefore the
decoder can be made open-loop and so unconditionally stable. We call this
design narrow-band, open-loop (NBOL).

\begin{figure}
\centerline{\psfig{figure=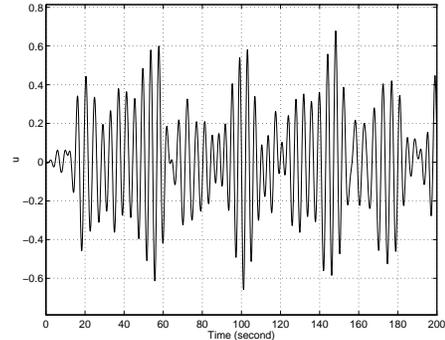,height=1.8in}}
\vspace{0.0in}
\caption{Time series of the output $u(t)$ from the system described
in Fig.\protect\ref{fig9}
\label{fig10}
}
\end{figure}

A natural approach to this end is to introduce a narrow bandpass filter in 
the feedback loops of both oscillators but eliminate compensating 
feedback loops (Figure 7b). In Eqs.(\ref{encoder}), (\ref{decoder}) 
this method corresponds to choosing $u_e\equiv \phi\otimes x_e + m(t)$. 
Such filtering however affects the
dynamics of the systems and can in principle destroy its chaotic 
attractor. In order to maintain chaos, one needs to have very strong 
nonlinearity (function $G(u)$) in the system, so
the narrow-band signal $u_e$ generates many spectral components
which in turn interact with other system variables. In numerical simulations,
we chose $G(u)=\sin(Au)$, and parameter $A$ characterizes the strength of 
nonlinearity. In our numerical simulations we used $A=15$. In the feedback 
loops fourth-order Butterworth bandpass 
filters are placed (center frequency 1.5 rad/sec, bandwidth $\Delta\omega=$
0.5 rad/sec). The chaotic attractor of this system in $(x,z)$ looks 
rather complicated (see Figure 9), but the variable $u$ at the output of the 
filter is indeed narrow-band (Figure 10). We coupled two such chaotic 
systems via a channel which is simulated by the  low-pass sixth-order
Chebyshev type II filter with cutoff frequency $\Omega_c=5$ rad/sec 
and 40 dB attenuation in the stopband. We transmitted two types of signals 
using this system. First, a sinusoidal signal $m(t)=0.1\sin(1.5t)$ which 
frequency coincides with the center frequency of feedback filters,
was injected in the encoder. The restored signal and the recovery error
are shown in Figure 11a,b. One can see that the signal recovery is
indeed very good. For comparison, we show in Figure 11c the recovered 
signal for this system without bandpass filters in the feedback loops. 
The ``recovered'' signal is not even close to the expected sine wave. 

\begin{figure}
\centerline{\psfig{figure=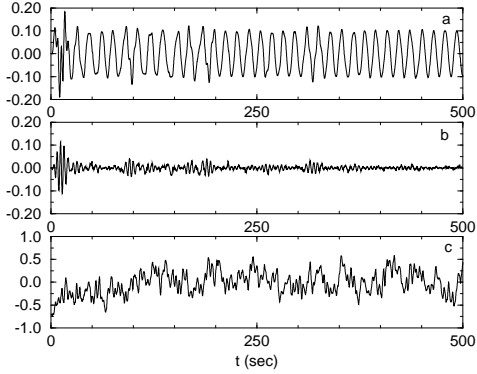,height=2.2in}}
\vspace{0.0in}
\caption{Recovered signal $m_r(t)$ {\em (a)} and recovery
error $m_r(t)-m(t)$ {\em (b)} for the system Fig.\protect\ref{fig7}b
with
chaos generator described in Fig.\protect\ref{fig9};
{\em (c)} recovered signal $m_r(t)$ for the same system with no bandpass
filters in feedback loops of chaos generators.
\label{fig11}
}
\end{figure}

\begin{figure}
\centerline{\psfig{figure=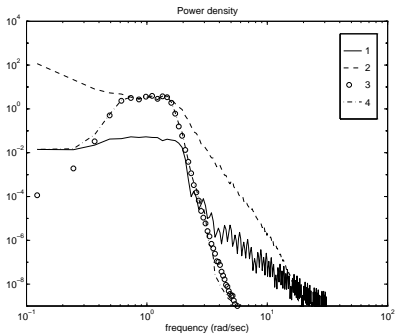,height=1.8in}}
\caption{Power spectra of the modulated binary sequence
(\protect\ref{pulseband}) {\em (1)},
$x$-component of encoder described in Fig.\protect\ref{fig9}{\em (2)},
transmitted signal $u_e$ {\em (3)}, and received signal $u_d$ {\em (5)}.
\label{fig12}
}
\end{figure}

We also carried a number of tests involving transmission 
a pseudo-random binary sequence $s_n$ of 1 and -1 using a pulse-band 
modulation 
\begin{equation}
m(t)=0.1\sum_{n=0}^{N}s_ng(t-nT)
\label{pulseband}
\end{equation}
for various filters in the feedback loops and in the channel.
Function $g(t)=\sin(1.5t)/(1.5t)$ was chosen to minimize intersymbol
interference at the demodulating stage, and $T=\pi/1.5$ sec. In Figure 12,
the power spectrum of the input signal $m(t)$ is shown together with
the power spectra of the $x$ component of the encoding system,
$u_e$ variable which is sent through the channel, and 
$u_d$ variable which is received at the decoder, for $\Omega_c$=4 rad/sec.
The time series of the 
modulated signal $m(t)$ is shown in Figure 13a. The recovered signal 
(starting from initial transient) at the decoder is shown in Figure 13b. 
After demodulating this signal we
computed error rate per 1000 bits for a number of values of $\Omega$ and
$\Delta\omega$. This data is presented in Figure 14. For comparison,
we also computed the error rate for the case when there is no filter
in the feedback loops of the systems. It is easy to see that
the using filters in the systems allows to reduce drastically the 
error rate once the signal $u$ fits into the channel passband.

\begin{figure}
\centerline{\psfig{figure=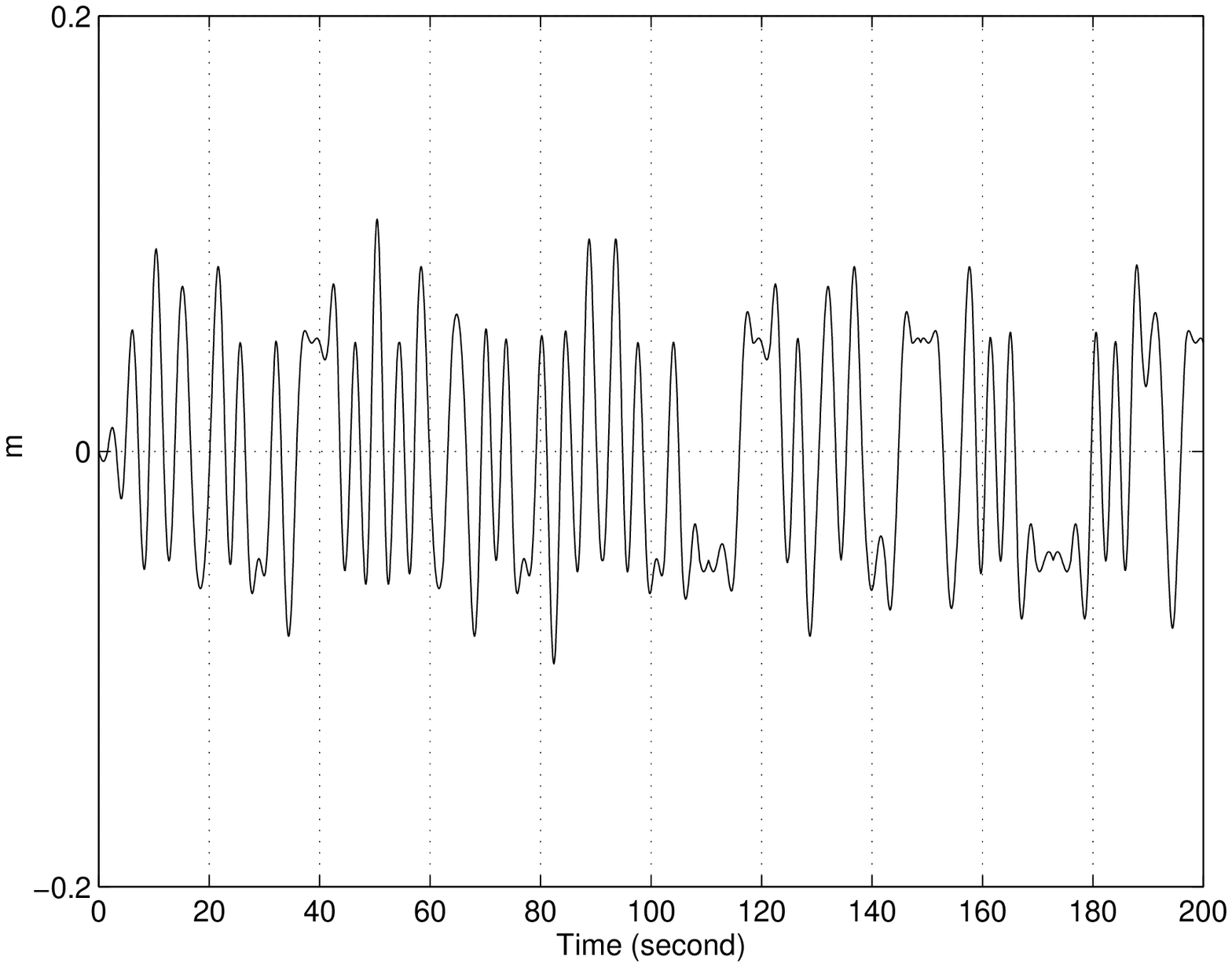,height=1.8in}}
\centerline{\psfig{figure=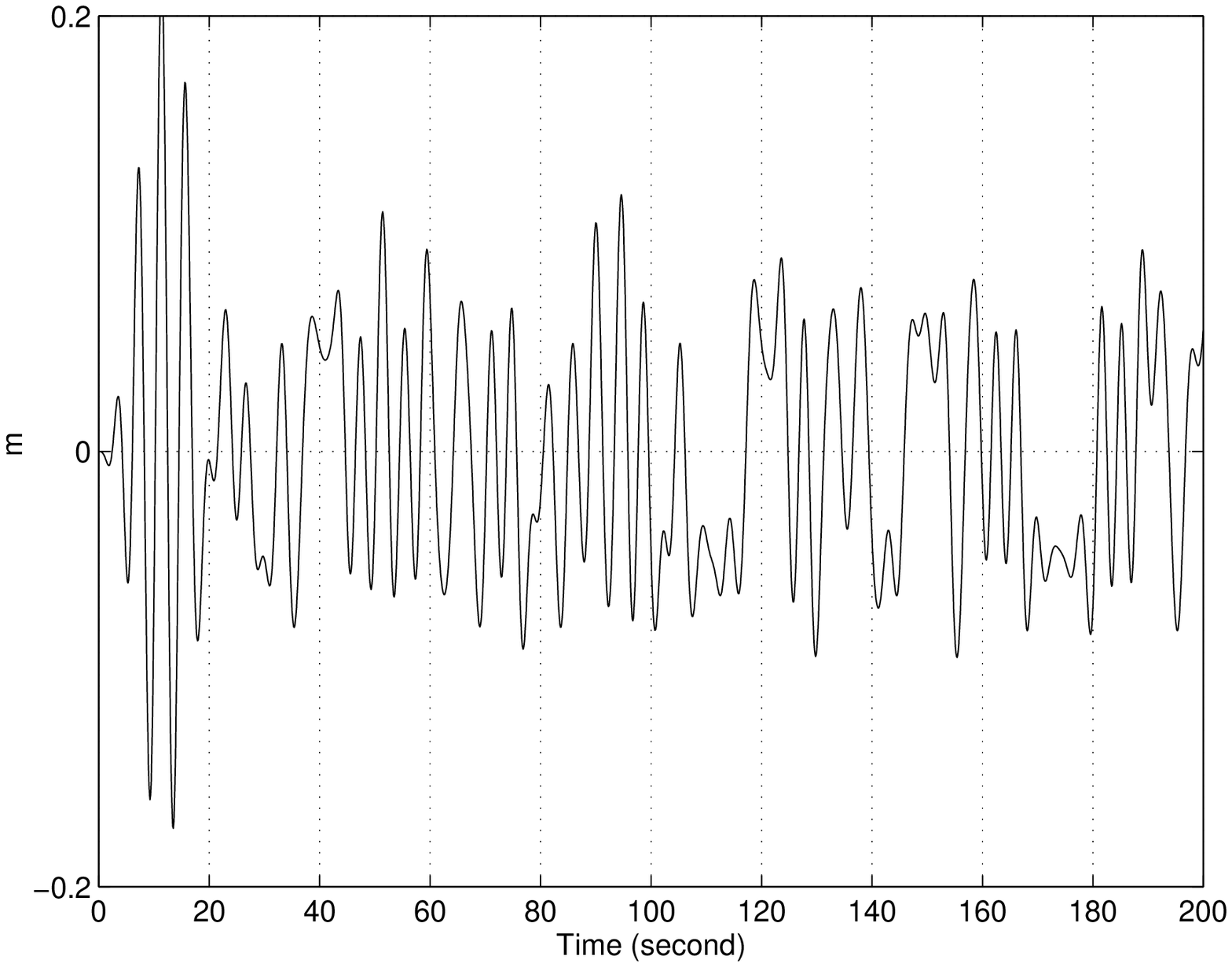,height=1.8in}}
\caption{Time series of original {\em (a)} and recovered {\em (b)}
modulated binary sequence (\protect\ref{pulseband}).
\label{fig13}
}
\end{figure}

\begin{figure}
\centerline{\psfig{figure=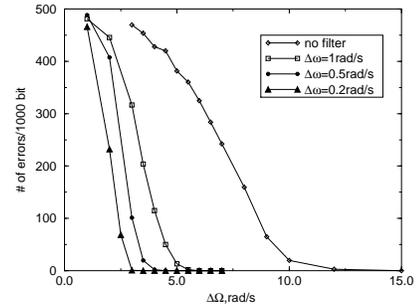,height=1.8in}}
\vspace{0.0in}
\caption{Error rate (per 1000 bits) for binary transmission
through the low-pass channel filter for various bandwidths of BPFs in
the encoder and decoder.
\label{fig14}
}
\end{figure}

\section{Experimental Results}
\label{sec5} 
 
As a generator of chaos we used an analog nonlinear circuit which in
its basic design consists of two input-output elements, nonlinear
converter (NLC) and a linear feedback (LFB), see Fig.~\ref{blkcirc}.
The details about implementation of NLC and LFB can be found in
the Appendix. Being connected in a loop, NLC and LFB give rise to 
self-sustained oscillations. By judicious selection of the nonlinearity
and parameters of the linear feedback one can achive
generation of chaotic oscillations. To design chaotic
encoder and decoder on the basis of this chaos generators we
use the method proposed in~\cite{volk93} (see Fig.\ref{fig5}a).  

\begin{figure}
\centerline{\psfig{figure=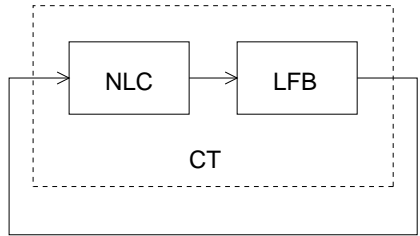,height=1.2in}}
\caption{Block-diagram of the nonlinear circuit used as
the generator of chaos. See Appendix for details on
NLC and LFB.
\label{blkcirc}
}
\end{figure}

\begin{figure}
\centerline{\psfig{figure=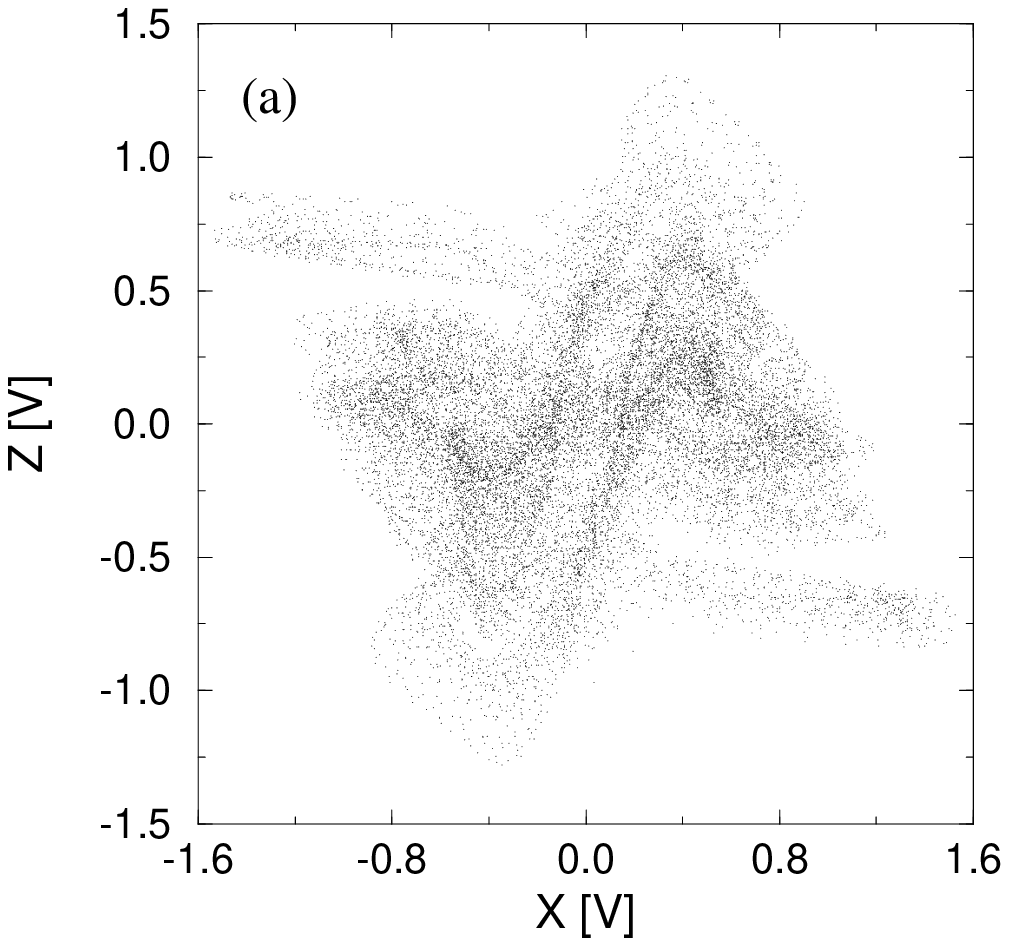,height=3.8in}
\hspace{-1.0in}\psfig{figure=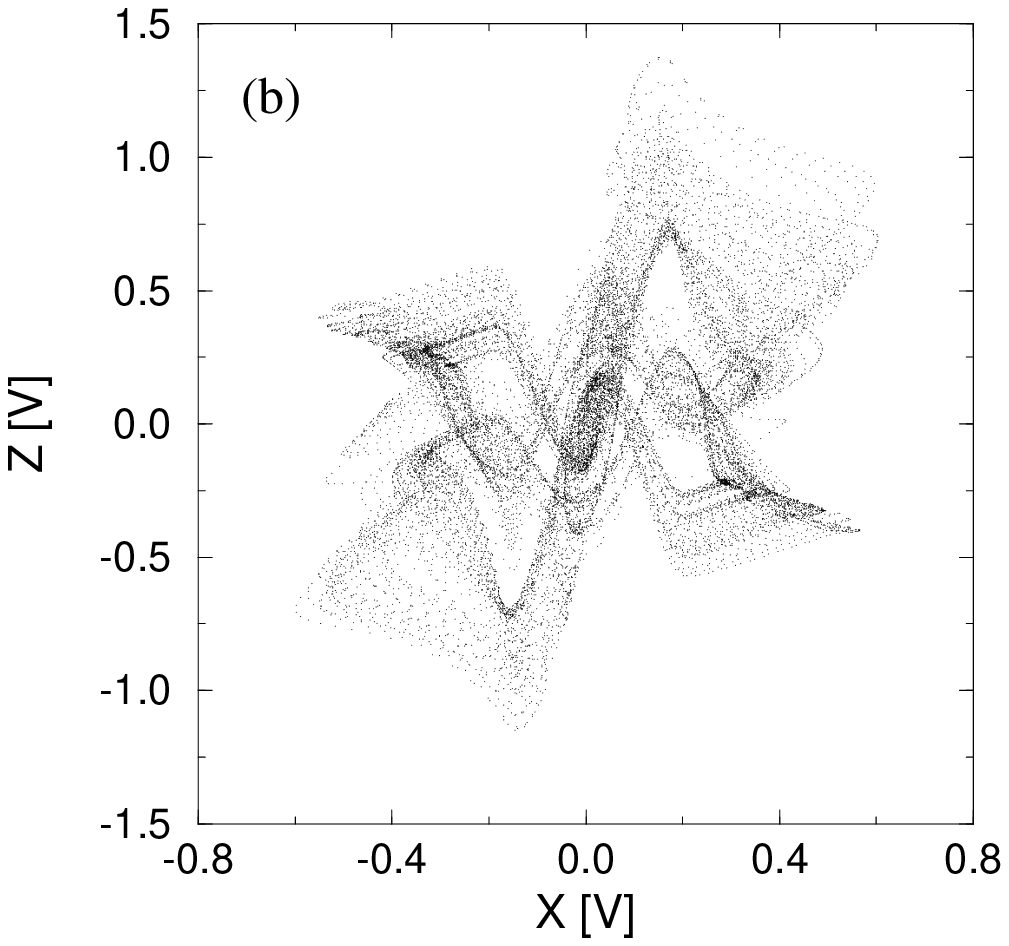,height=3.8in}}
\vspace{-1.8in}
\caption{Chaotic attractors generated by the encoder
with $m(t)=0$. {\em (a)} Original encoder, $R_n$=1103~Ohm.
{\em (b)} Modified narrow-band encoder, $R_n$=446~Ohm.
\label{attractors}
}
\end{figure}

To model limited bandwidth of the communication channel we
connected the encoder and decoder through a second-order low-pass
VCVS active filter, see Appendix for details. In the diagrams
Fig~\ref{fig5} this filter is labeled CHF. Parameters
of the filter were selected to make a Bessel filter~\cite{horowitz}.
As a control parameter of the filter we use its cutoff frequency, 
$f_c$ measured at -3dB.
 
In the experiments we compared two schemes of chaos
communication. The first is the original scheme
implemented with open-loop chaotic decoder, see Fig~\ref{fig5}a. The second
one is the modified scheme Fig.\ref{fig7}b that  
utilizes low-pass filters (LPF) to design the narrow bandwidth signal 
in the feedback of the chaos generator. To recover the message in the
modified scheme we used a matched narrow-band open-loop (NBOL) chaotic 
decoder, see Fig.\ref{fig7}b. The diagram of LPF is presented in the 
Appendix. Parameters of the filters were selected to make Chebyshev filter
with the cutoff frequency $f_\phi$=300Hz measured at -2dB.
Chaotic attractors generated by the chaotic encoders of
the original and the modified schemes with $m(t)=0$ are shown
in Fig.~\ref{attractors} a and b, respectively. Variable $x(t)$
is the voltage measured at the output of LFB, and variable $z(t)$
is the voltage across the capacitor C1 in the LFB, see Appendix A.
 
\begin{figure}
\centerline{\psfig{figure=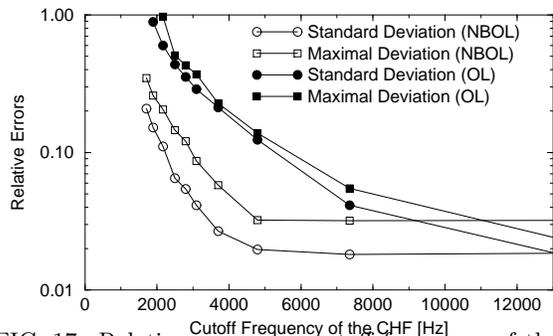,height=2.5in}}
\vspace{-0.8in}
\caption{Relative errors vs. cutoff frequency of the filter in
the communication link in the experiment with electronic circuits.
\label{errors}
}
\end{figure}

The main goal of the experiment was to measure the
quality of synchronization as a function of the cutoff frequency
of filter CHF. For this purpose,
it is expedient to apply $m(t)=0$, and measure $u(t)$ and $m_r(t)$. 
Using these waveforms sampled at 
10 KHz we calculated relative values of the maximal deviation, 
$\mbox{max}|m_r(t)|/\mbox{max}|u(t)|$, and standard deviation
\begin{equation}
\frac{<m_r^2-<m_r>^2>^{1/2}}{<u^2-<u>^2>^{1/2}}
\end{equation}
Both these characteristics are normalized with respect to the
parameters of the driving signals. These errors as functions of 
the cutoff frequency of the channel are
shown in Figure~\ref{errors}. The plots clearly how demonstrate that
using additional filters in the encoder and decoder
can significantly reduce errors caused by
degrading the chaotic signal in the limitied bandwidth channel
In the case of NBOL errors becomes significant only when the channel 
has the cutoff frequency less than 4~kHz. For the broader channel the errors
are due to the slight mismatch between the parameters of the encoder
and decoder.
 
Since in the considered method of chaos communication
the message $m(t)$ does not distroy the symmetry between 
encoder and decoder, the results of error evaluation can also be
applied, to some extent, to assess the quality of communication 
($m(t)\neq0$). However, when the width of the spectrum of the 
information signal becomes comparable with the bandwidth of the 
channel the errors will significantly increase. 

\begin{figure}
\centerline{\psfig{figure=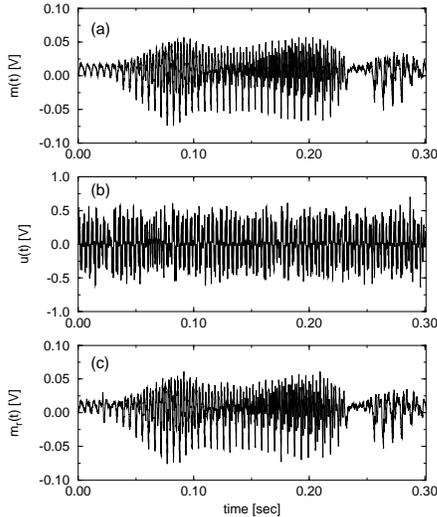,height=3.3in}}
\vspace{-0.7in}
\caption{Waveforms of the original information signal {\em (a)},
the encoded signal measured in the communication link {\em (b)}, and
the recovered information signal {\em (c)}. The cutoff frequency
of the CHF 4.8~kHz, sampling rate 30000 samples per second.
\label{commun}
}
\end{figure}

To demonstrate the feasibility of the communication with
NBOL through the narrow bandwidth channel we used voice signal 
as $m(t)$ and measured $m(t)$, $u(t)$ and $m_r(t)$
simultaneously. Figure~\ref{commun} shows the results
of the measurements obtained for the CHF with $f_{c}=
4.8~kHz$. 
Although
one can notice imperfections of the recovered signal, the quality 
of the communication is pretty
good. Note that for this cutoff frequency of the channel filter
the standard scheme (without LPF) does not yield any meaningful 
signal recovery (cf. Fig.~\ref{errors}).

\section{Conclusions}
In this work we have explored various possibilities of transmitting
analog information masked by a chaotic signal through a band-limited channel.
Two main classes of systems are considered. First (adaptive) class
corrects the channel distortions either using inverse filter
at the receiver, or augmenting the transmitted
signal by the chaotic signal generated by the decoding chaotic system
passed through a complimentary matched filter. Both these approaches
suffer from potential instabilities leading to the poor signal recovery.

The second class is based on modification of the structure 
of the chaotic signal {\em before} sending it through the channel, so it 
is by construction made suitable for undistorted transmission. This
approach does not require learning the channel characteristic and 
can provide robust signal transmission. We demonstrated in numerical 
simulations and experiments with electronic circuits that the quality of
signal recovery improves greatly as compared to transmission with 
wide-spectrum chaotic signals.

\section{Acknowledgments}

We are grateful to H.Abarbanel and M.M.Sushchik for 
helpful discussions.  This work was supported in part by the 
U.S. Department of Energy under contract DE-FG03-95ER14516, and 
in part by the U.S.Department of the Air Force (SBIR, contract 
F19628-96-C0076).

\section{Appendix: Elements of the Experimental Setup}
 
This Appendix contains the diagrams of the circuits
used to build the chaos generator, auxiliary filters
and the model of band-limited communication link.
 
To build the generator of chaos we used the loop type
oscillator design~\cite{dmitriev}. This design employs
serial connection of input-output stable elements.
When such elements are connected into the loop, see Fig~\ref{blkcirc}
the whole system can become a generator of self-sustained
oscillations. By selecting characteristics of
the input-output elements one can always achive the regime
of chaos generation. In our experiments we used two input-output
elements: nonlinear converter (NLC) and linear system
which forms a linear feedback (LFB) of the oscillator.

\begin{figure}
\centerline{\psfig{figure=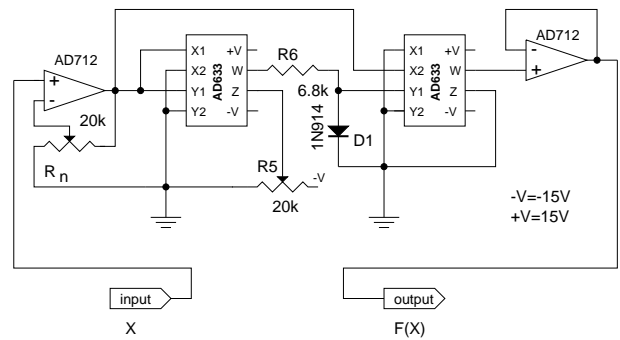,height=1.7in}}
\caption{Diagram of the nonlinear converter (NLC)
\label{nlc}
}
\end{figure}

\begin{figure}
\centerline{\psfig{figure=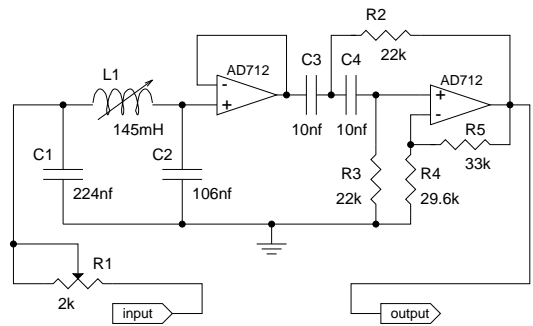,height=1.7in}}
\caption{Diagram of the linear feedback element (LFB)
\label{lfb}
}
\end{figure}

\begin{figure}
\centerline{\psfig{figure=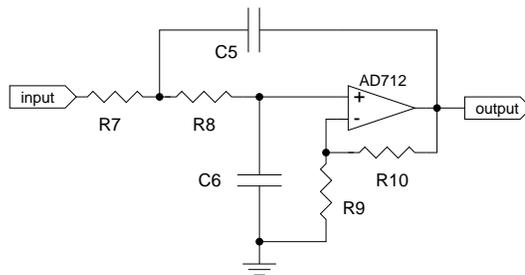,height=1.4in}}
\caption{Diagram of the second order low-pass VCVS filter.
\label{lpf}
}
\end{figure}

The diagram of the NLC is shown in Fig.~\ref{nlc}. Symmetric
nonlinear characteristics of the converter are implemented with
two multipliers AD633 and diod D1. Resistor R1 is used
to supply 9.3V at the offset Z of the first multiplier. 
Variable resistor $R_n$ was used to select the desired regime
of oscillations ($R_n$ specifies the value of resistance between 
the ground and $-IN$ terminal of the OP-amplifier.
 
The circuitry of the linear part of the chaotic oscillator is
shown in Fig.~\ref{lfb}. It uses RC and resonant circuits to create
sufficient number of degrees of freedom (dimension of the phase space)
required for chaotic dynamics. In addition to that, LFB supplies 
a second order
high-pass filter. This filter is used to eliminate
DC components of the chaotic signal which is used as
communication signal. In experiments the value of the
resistor R1 was set at 1.01~KOhm.
 
As the low-pass filters in the modified chaos generator and 
the filter in the model of communication channel we used the
second order filter which diagram is shown in Fig.~\ref{lpf}.
For LPF we used the filter with the following parameters:
C5=C6=95.4~nF, R7=R8=6.11~KOhm, R9=24.7~KOhm, and R10=27.6~KOhm.
For CHF we set R7=R8=22.3~KOhm, R9=50~KOhm, and R10=13.4~KOhm.
The cutoff frequency was selected by capacitors C5=C6. Summation and
subtraction were implemented using standard schemes of summing and
differential amplifiers\cite{horowitz}.

\end{document}